\documentclass[a4paper,10pt]{article}
\pdfoutput=1
\usepackage[top=2.5cm, bottom=2.5cm, left=2.0cm, right=2.0cm,
columnsep=0.8cm]{geometry}
\usepackage{enumitem}
\usepackage[hidelinks]{hyperref}
\usepackage{boxedminipage}
\usepackage{nopageno}
\usepackage{graphicx}
\usepackage[numbers]{natbib}
\usepackage[font=it]{caption}
\usepackage[usenames,dvipsnames]{xcolor}
\usepackage{listings}
\usepackage{caption}
\usepackage{subcaption}
\usepackage{todonotes}
\usepackage{color,soul}

\usepackage{lastpage}
\usepackage{fancyhdr}
\setlist{itemsep=-0.1cm,topsep=0.1cm,labelsep=0.3cm}
\setenumerate{leftmargin=*}
\makeatletter
\renewcommand\title[1]{\gdef\@title{\fontsize{12pt}{2pt}\bfseries{#1}}}
\makeatletter
\renewcommand\section{\@startsection{section}{1}{\z@}{3pt}{3pt}{\normalfont\large\bfseries}}
\makeatletter
\renewcommand\subsection{\@startsection{subsection}{1}{\z@}{\z@}{\z@}{\normalfont\normalsize\bfseries}}
\makeatletter
\renewcommand\subsection{\@startsection{subsection}{1}{\z@}{\z@}{0.1pt}{\normalfont\normalsize\bfseries}}

\newcommand{\greyBoxedMinipage}[1]{\fcolorbox{lightgray}{lightgray}{ \begin{minipage}{0.75\textwidth} #1 \end{minipage}}
}

\title{Method development for lowering supply temperatures in existing buildings \\ \vspace{4pt} using minimal building information and demand measurement data}

\author{
Jan Stock$^1$, Philipp Althaus$^1$, Sascha Johnen$^1$, \\ André Xhonneux$^1$, Dirk Müller $^{1,2}$\\ 																										
$^1$ Forschungszentrum Jülich GmbH, Institute of Energy and Climate Research, \\ Energy Systems Engineering
(IEK-10), 52425 Juelich, Germany\\ 																																	
$^2$ E.ON Energy Research Center, Institute for Energy Efficient Buildings and Indoor Climate, \\
RWTH Aachen University, 52056 Aachen, Germany\\ 																																	
\phantom{Line 9}
} 																																									
\date{\vspace{-0.5cm}}	
\begin{document}

\maketitle
\begin{center}
	\greyBoxedMinipage{This work is successfully peer-reviewed and accepted at BuildingSimulation 2023.} 
\end{center}

\section*{Abstract}
\addtocounter{section}{1}

Regarding climate change, the need to reduce greenhouse gas emissions is well-known. As building heating contributes to a high share of total energy consumption, which relies mainly on fossil energy sources, improving heating efficiency is promising to consider.
Lowering supply temperatures of the heating systems in buildings offers a huge potential for efficiency improvements since different heat supply technologies, such as heat pumps or district heating, benefit from low supply temperatures.
However, most estimations of possible temperature reductions in existing buildings are based on available measurement data on room level or detailed building information about the building's physics to develop simulation models.

To reveal the potential of temperature reduction for several buildings and strive for a wide applicability, the presented method focuses on estimations for temperature reduction in existing buildings with limited input data. 
By evaluating historic heat demand data on the building level, outdoor temperatures and information about installed heaters, the minimal actual necessary supply temperature is calculated for each heater in the building using the LMTD approach. 
Based on the calculated required supply temperatures for each room at different outdoor temperatures, the overall necessary supply temperatures to be provided to the building are chosen. Thus, the minimal heatcurve possible for an existing building is deduced.

The method described is applied to multiple existing office buildings at the campus of Forschungszentrum Jülich, Germany, demonstrating the fast application for several buildings with limited expenditure. 
Furthermore, a developed adapted heatcurve is implemented in one real building and evaluated in relation to the previously applied heatcurve of the heating system.
\section*{Highlights}
\begin{itemize}
\item Identification of supply temperature reduction potential with limited input data
\item Supply temperature estimation based on heat consumption data and heater information
\item Room heat load estimation based on basic building geometry information and heat transfer coefficients of typical buildings
\item Clustering of measurement data shows various heat consumption patterns that lead to heatcurve distinction
\end{itemize}
\section*{Introduction}

Reducing carbon dioxide emissions in existing energy infrastructures is crucial for a sustainable energy transition. In particular, the existing building stock offers enormous potential for reducing carbon dioxide emissions, which are mainly related to heat supply.
Retrofitting existing buildings can reduce heat demand through the refurbishment of the buildings' envelope or increase heat supply efficiency by installing modern heating systems, but it also involves high expenditures.
Alternatively, adjusting the operating temperature of the heating system is an opportunity to increase the supply efficiency in existing buildings and, thus, reduce carbon dioxide emissions.
Lowering the heat curve of a heating system, which controls the supply temperature setpoint in the building, is solely an operational adjustment with little effort compared to retrofitting the building, but it still has multiple positive effects on the building's heat supply. 

The supply temperature level of the heating system plays an important role in the efficiency of different heat generation technologies. 
Condensing boilers operate most efficiently with minimal return temperatures to utilise the latent heat of the condensation process (\citet{Satyavada.2016}
). Minimal return temperatures at the condensing boiler can be achieved by lowering the heat curve of the heating system. 

Moreover, heat pumps and district heating systems also benefit from low supply temperatures (\citet{Hesaraki.2015}).
For instance, the efficiency of heat pump operation depends on the temperature difference between the heat source and the heat sink, i.e. the heating system of a building. Most used renewable heat sources by heat pumps in buildings are at a low temperature level, such as ambient air or shallow geothermal sources. 
Huchtemann et al. present the positive effects of supply temperature reduction for an air-water heat pump in a simulation study with an adapted control algorithm that considers the actual heat demand (\citet{Huchtemann.2013}).

Lowering the temperature requirements in the heating systems of buildings also has positive effects on district heating systems. By lowering the required supply temperature of the supplied buildings, the overall operating temperature of the district heating system can also be reduced, resulting in a reduction of heat distribution losses. Furthermore, the central heat supply benefits from reduced temperatures, e.g., heat pumps and condensing boilers are used as heat plants for district heating (\citet{Lund.2018}). 
In addition, several non-fossil heat sources can be utilised much easier in existing district heating systems if the operating temperature is lowered, e.g. most waste heat sources are available at a relatively low-temperature level compared to current average district heating supply temperatures (\citet{Jodeiri.2022}).

When adapting the heat curve of a heating system within an existing building, the minimum temperature requirements to meet the heat demand of the users in the building must be obtained. Therefore, the technical feasibility of sufficient heat transfer via the installed heaters in rooms with lowered temperatures must be considered. 
Since older radiators were often small in size to save costs, they require high supply temperatures of around 90/70~°C (design supply/return temperature), while newer installed heaters in modern buildings can already manage with lower temperatures around 55/35~°C (\citet{stergaard.2022}). 

Most heaters in existing buildings can be operated with lower supply temperatures than their design temperature indicates due to a decreased heat demand of the building compared to the heat demand at construction time or due to the fact that the installed heaters were initially oversized (\citet{Reguis.2021}). 
The potential of using the initial oversizing is investigated by \citet{stergaard.2016b} as they study the required supply temperatures of heating systems in older buildings in Denmark and show potential for operation with lower supply temperatures than specified by the heater design. 
Furthermore, small measures at the heating system can exploit additional potentials for temperature reduction, such as replacing critical heaters or installing local control devices (\citet{stergaard.2022}).
In \citet{Brand.2013}, the temperature reduction possibilities for different building renovation stages are investigated. They show that even small renovations, such as the replacement of windows, can reduce the supply temperature from an initial 78°C to under 60°C for most of the year, and extended renovations could reduce the temperature to 50°C.
However, renovations of the building envelope and retrofits of the heating systems are associated with high costs. Therefore, the implementation in a whole district to extend the full effects, for instance, for district heating supply, is limited.

Tunzi et al. present a method for optimising the temperature reduction in a building without retrofitting the building envelope or heating system by using the logarithmic mean temperature difference (LMTD) method, which allows a heat supply estimation of heaters based on temperature adaption and valve control adjustments (\citet{Tunzi.2016}).
Benakopoulos et al. evaluate data from heat allocators and show that not every installed heater in a building is used for heat supply and that temperature reduction is possible if every heater is used in a part load operation (\citet{Benakopoulos.2021}).
In \citet{Benakopoulos.2022}, the actual required supply temperature at the heaters is calculated using the measurement data of the installed heat allocators and the calculated heat demand of the building. 
The presented calculations to determine temperature reduction potentials in existing buildings are based on dynamic simulation models or the data availability of heat allocators at room level. The development of extensive simulation models for each investigated building or the installation of several heat allocators is associated with high effort and requires specialist knowledge. Thus, an affordable scaling for a broad application on the district level is limited.

To boost the approach of exploiting available potentials of lower supply temperatures and the associated efficiency improvements, we present a method based on minimal data input to enable a wide application range to multiple buildings with limited effort.
Therefore, we develop an approach to estimate the heat load of each room based on heat consumption measurements on the building level and compare it with the capacities of the installed radiator heaters. Based on the known heater capacities and the estimated demand of each room, the actual required operating temperatures of each heater can be derived resulting in an adapted heatcurve for the total radiator-based heating system. 
Thus, the proposed method does not rely on changes in the installed heating technology or the general control approach of it.

Following the introduction and literature review in this section, we present our developed method in the following section. 
Thereafter, we show our reduced heatcurves results and present the experimental application of an adapted heatcurve at one example building. After the discussion of the proposed method and presented results, we conclude our investigation.
\section*{Methodology}
\label{sec:Method}

For a broad application to existing buildings, we develop a method based on minimal data input. The required data should be available with little effort. 
The proposed method combines static building information and measured data to derive a heatcurve adapted to the actual operation of the radiator-based heating system. The steps to conduct are the following: 

The \textbf{first step} is the data collection of static building information, including floor plans and heater characteristics, statistical heat transfer coefficient of typical building types and historic trajectories for outdoor temperature and heating demand of the building.

In the \textbf{second step}, a model is developed to derive the building consumption $\dot{Q}_{mod}$ dependent on time of day and outdoor temperature $T_{out}$ from the measurement data.
The time dependency is evaluated by assigning each ten minute interval of a day to one of $n_{cluster}$ using the kmeans algorithm. This allows to identify regions of different dynamic behaviour, such as reduced heat loads during the night or times with high demand at morning. The features used as input for the cluster algorithm are the mean demand as well as 90\%- and 10\%-quantile for each 10 minute interval, as commonly used in statistics.
Since each cluster should resemble a region of different behaviour the temperature dependency is modelled per cluster. Using the maximum value of heat consumption per outdoor temperature $T_{out}$ would likely result in artificially increased heat loads, so the 90\%-quantile for each interval is used.
All following steps are calculated per cluster and outdoor temperature to provide heatcurve values for each time frame identified using the clusters. 

Within \textbf{step three}, the share of the $i$-th room heat load $\dot{Q}_{i,mod}$ on the total building consumption is estimated using typical heat transfer coefficients, i.e. U-values, for walls (wa), windows (wi), roofs (ro) and floor slab (fs), as well as data about the area of each room's boundaries causing an energy transfer to the environment.
To determine the heat load of each room based on these information, a system of equations is developed. The system is first filled with equations for the unknown heat losses of each room, i.e. the required heat load of each room, $\dot{Q}_{i,mod}$ 

\begin{equation}
\label{eq:heat_loss}
    \dot{Q}_{i,mod}(T_{out}) = \sum\limits_{k=1}^{n_{boundaries}}A_{i,k} \cdot U_{k} \cdot (T_{in} - T_{out}),
\end{equation}

where the room's heat load is expressed as the sum of all individual heat flows through the building envelope, described by corresponding heat transfer coefficient $U_k$ and the surface area of the adjacent part of the room's envelope $A_{i,k}$, and the temperature difference between the indoor temperature $T_{in}$ and $T_{out}$. The corresponding value of $T_{out}$ is taken from the clustered measurement data and the size of adjacent surfaces is calculated with information from floor plans. In case of $T_{in}<T_{out}$ no equation is added, to avoid negative heat loads.
Since the specific values $U_k$ for the various adjacent building parts are often unknown, additional correlations are needed to replace the heat transfer coefficient. Therefore, ratios of typical heat transfer coefficients ($U_{ratio} = U_{l} / U_{m}$) of typical buildings are inserted to the system of equations. Based on the year of construction and the building type of the investigated building, values of heat transfer coefficients are taken from a building category of the German building stock according to TABULA (\citet{Loga.2015}). Thus, any three ratio correlations are added to the system of equations deduced of all possible combinations for the four adjacent building parts taken into consideration (wa, wi, fs, ro).

\begin{equation}
\label{eq:U_ratio}
    U_{ratio,1-3} = \frac{U_{l}}{U_{m}}; \quad  l, m \in \{wa, wi, fs, ro\}
\end{equation}

By using this approach, no absolute values for the heat transfer coefficients are required, as the actual values still differ between individual buildings. Since the total heat consumption of the building is known, the ratios are used to estimate the relative heat losses depending on the proportion of adjacent building parts. 
The heat transfer to adjacent rooms with different indoor temperature setpoints is neglected.

To solve the system of equation and, thus, to determine the absolute heat load of each room, the total building heat demand $\dot{Q}_{mod}$ at a given $T_{out}$ is expressed by the sum of all heat loads of each rooms and added to the system as final equations.

\begin{equation}
\label{eq:Q_i,mod}
    \dot{Q}_{mod}(T_{out}) = \sum\limits_{i=1}^{n_{rooms}}\dot{Q}_{i,mod}(T_{out})
\end{equation}

An analysis of the heat loads in the hallways indicates that the heaters are often undersized to meet the calculated heat loads, which is in reality compensated by the additional supply through the adjacent rooms. To consider possible undersizing, the heaters capacities in hallways are estimated on the basis of a highly reduced heat curve, which is assumed to be lower than the possible supply temperature reduction. The assumed temperature for the hallways heaters is afterwards verified to be not the limiting temperature for the overall temperature reduction. The remaining heat load of the hallway, not met by the assumed and reduced heater capacity compared to nominally values, is partitioned to the adjacent rooms, based on the ratio of corresponding wall surfaces.
The staircases are compromised into one room, as there are no separating elements between the floors in the investigated buildings.

\textbf{Step four} uses the found heat load per room to derive the required LMTD per installed heater $h$. In general the heatflux provided by a single heater can be described by (\ref{eq:LMTD_Q})
where $A_h$ is the heat exchanger surface of the heater with the heat transfer coefficient $U_h$ and a constant $n$ which is dependent on the heater type. 
The LMTD is calculated by (\ref{eq:LMTD}) with the supply $T_{sup}$ and return $T_{ret}$ temperature at the heater.

\begin{equation}\label{eq:LMTD_Q}
    \dot{Q}_{h} = U_h \cdot A_h \cdot \left( LMTD\right)^n
\end{equation}

\begin{equation}\label{eq:LMTD}
    LMTD = \frac{T_{sup} - T_{ret}}{\ln{\frac{T_{sup} - T_{in}}{T_{ret} - T_{in}}}}
\end{equation}

The LMTD is calculated with (\ref{eq:LMTD}) for each heater $h$. Once at nominal values based on data sheets $\dot{Q}_{h,nom}$, and once for the required heat load of each room $\dot{Q}_{i,mod}$ according to the different outer temperatures. 
In case of one heater installed per room, the room heat load $\dot{Q}_{i,mod}$ is the heat load of heater $\dot{Q}_{h,mod}$. In case of multiple heaters per room, the room heat load  $\dot{Q}_{i,mod}$ is split to the number of heaters, to gain the required heat load of each heater $\dot{Q}_{h,mod}$. 

The actual required $LMTD_{h,mod}$ for each heater is derived, by using the relation $\dot{Q}_{h,mod}/\dot{Q}_{h,nom}$, which results in


\begin{equation}\label{eq:LMTD_frac}
\small
LMTD_{h,mod} =  
    \left(\frac{\dot{Q}_{h,mod}}{\dot{Q}_{h,nom}}\right)^{\frac{1}{n}} \cdot LMTD_{h,nom},
\end{equation}

assuming that the temperature difference between supply and return $\Delta T_{h}=T_{sup} - T_{ret}$ for both LMTD is the same. 

The following \textbf{step five} deduces the minimum supply temperature needed per heater with (\ref{eq:LMTD2T:sup:nec}),  based on $LMTD_{h,mod}$ (\ref{eq:LMTD_frac}) and LMTD formulation (\ref{eq:LMTD}), to allow for supply the requested heat load for the room. 

\begin{equation}
\label{eq:LMTD2T:sup:nec}
T_{h,mod,sup} = \frac{T_{in} - e^{\frac{\Delta T_{h}}{LMTD_{h,mod}}} \cdot (\Delta T_{h} + T_{in})}{1 - e^{\frac{\Delta T_{h}}{LMTD_{h,mod}}}}
\end{equation}

Having calculated the necessary supply temperature per heater $T_{h,mod,sup}$, the necessary minimum supply temperature of the whole building $T_{mod,sup}$ is derived in \textbf{step six}. This is done by finding the maximum of the minimal necessary temperature per cluster and outside temperature (\ref{eq:TVL:min:building}). 

\begin{equation}
\label{eq:TVL:min:building}
T_{mod,sup} = max(T_{h,mod,sup}) 
\end{equation}

Thus, an adapted heatcurve, depending on the outside temperature, is deduced for each cluster. 
An offset for the set supply temperature at the heating system could be applied as safety margin to consider e.g. temperature losses. 

Finally, in \textbf{step seven}, the calculated heatcurve is postprocessed by extrapolating the curve where no weather or historical heat demand data is available. The values are extrapolated using front or back fill depending on whether they are below or above the reduced heatcurve. Finally the well-known savitzky-golay-filter is applied for smoothing the heatcurve. 

\section*{Results}
\label{sec:Result}
\subsection*{Heatcurve calculation}

Calculations according to the proposed method have been executed for several buildings. Within the Living Lab Energy Campus (LLEC) project, several existing buildings on the campus of the Forschungszentrum Jülich, Germany, are being equipped with sensors and actuators to enable monitoring of buildings' operation and the application of innovative control approaches (\citet{Althaus.2022}). Three of these buildings, which are part of the LLEC project, are selected to test the presented method. 

The basic information on building characteristics, number of chosen clusters, time spans for available measurement data and additional used hyperparameters for method calculation is given in Table~\ref{tab:BuildingOverview}. The building construction type is needed to derive typical U-values from TABULA (\citet{Loga.2015}).
Weather data has been used as provided by the German meteorological service (\citet{DeutscherWetterdienst.2022}).
The indoor temperatures in most rooms assumed to be 20~°C. Certain rooms, like bathrooms will be set to have a reduced value of 18~°C.

\begin{table}[ht]
\vspace{-5pt}   
\caption{Building characteristics and applied method hyperparameters.}
\label{tab:BuildingOverview}
\small
\centering
\begin{tabular}{| c | c | c | c |}
\hline
     \bf{Building} &
      \bf{A} &
      \bf{B} &
      \bf{C}
      \\
     
      \hline
    Constr. year &
      1976 &
      2021 &
      2011
      \\
      \hline
    Constr. type &
      MFH\_F &
      MFH\_J &
      MFH\_J
      \\
      \hline
    \# Rooms &
      60 &
      96 &
      70
      \\
      \hline
    \# Floors &
      2 &
      3 &
      3
      \\
      \hline
    Data: start &
       01/01/2021 &
       27/09/2022 &
       01/01/2021
      \\
      \hline
    Data: end &
       02/01/2022 &
       14/12/2022 &
       02/01/2022
      \\
      \hline
    Data: $T_{out}$ &
      -11 to 34~°C &
      -7 to 23~°C &
      -11 to 31~°C
      \\
      \hline
    \# Cluster &
      1 &
      3 &
      2
      \\
      \hline

\end{tabular}
\vspace{-5pt}   
\end{table}

\begin{figure}[htbp]
\centering
  \includegraphics[width=0.7\linewidth]{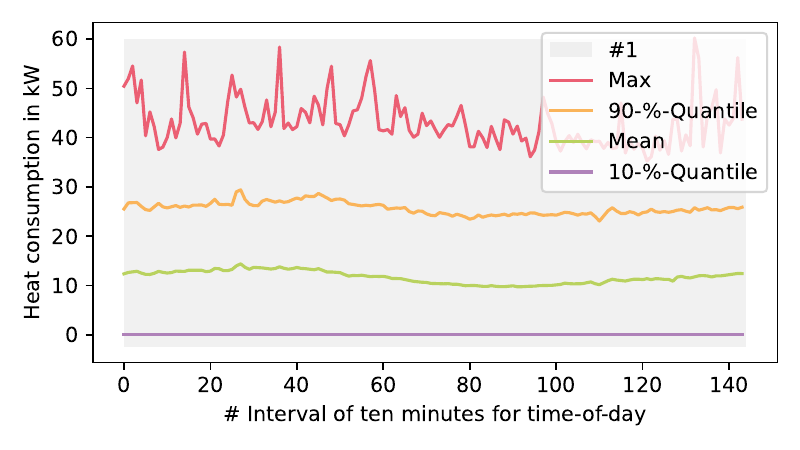}
  \caption{Quantiles of heat consumption data over ten-minute-intervals for building A.}
  \label{fig:Quantiles:buildA}
\vspace{-12pt}   
\end{figure}

Analysis of the heat consumption data over time-of-day for building A showed relatively uniform consumption. Since the clusters are intended to be used to identify regions of differing demand, only one cluster is used in this case. The heat consumption analysis of building A is shown in Figure~\ref{fig:Quantiles:buildA}.

\begin{figure}[htbp]
\centering
  \includegraphics[width=0.7\linewidth]{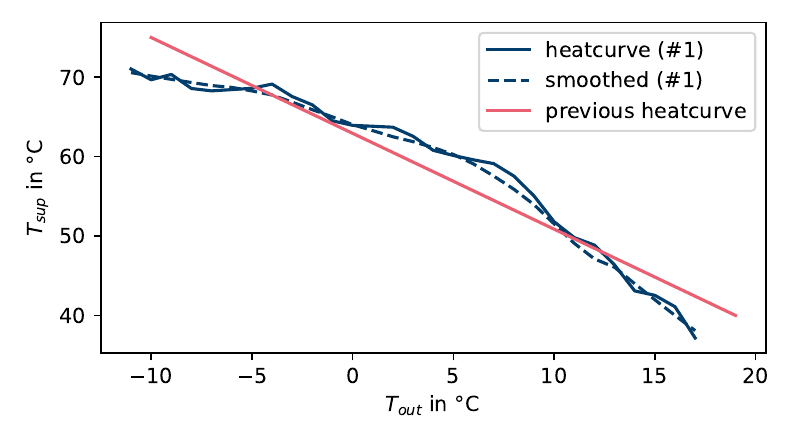}
  \caption{Calculated and previously active heatcurve for building A.}
  \label{fig:heatcurve:comparison:buildA}
\vspace{-12pt}   
\end{figure}

The resulting heatcurve for building A in comparison to the previously applied heatcurve during the timespan used for demand data sampling is shown in Figure~\ref{fig:heatcurve:comparison:buildA}. 
It becomes visible, that the calculated supply temperature slightly differs compared to the previously active one. At 5~°C outside temperature, the calculated heatcurve is ca. 3.4~K higher while being ca. 4.8~K lower at $-10$~°C outside temperature.

\begin{figure}[htbp]
\centering
  \includegraphics[width=0.7\linewidth]{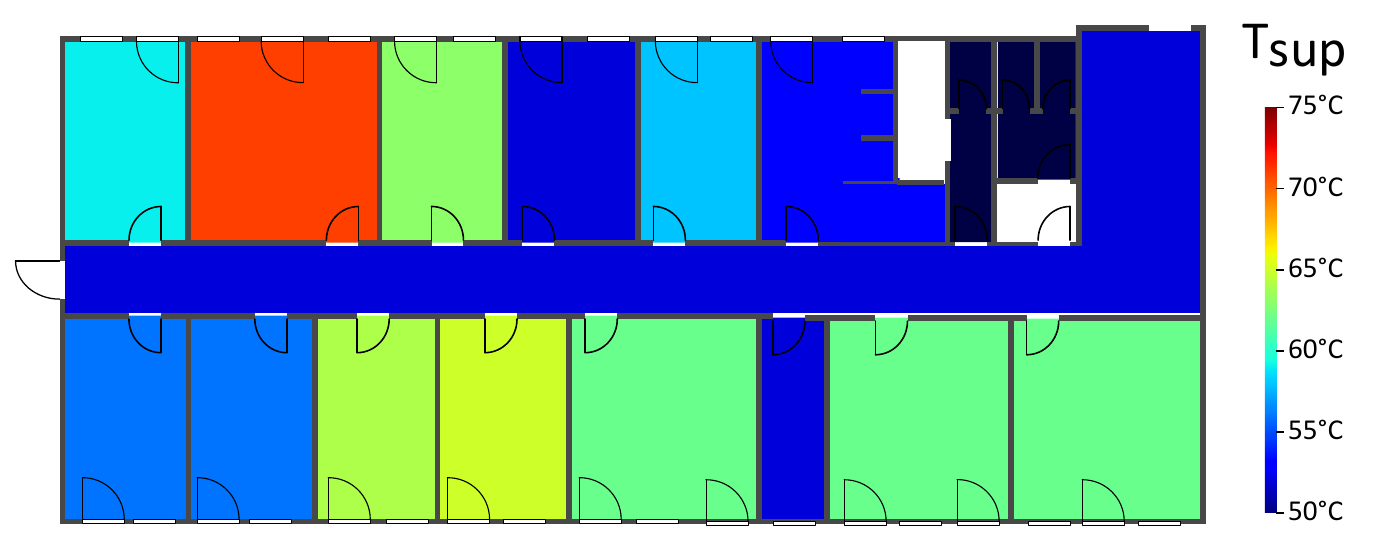}
  \caption{Excerpt of floor plan of building A containing the highest necessary heater supply temperature per room.}
  \label{fig:floor_plan:buildA}
\vspace{-12pt}   
\end{figure}

In Figure~\ref{fig:floor_plan:buildA}, the top floor of building A is shown, where the highest necessary heater supply temperature is identified. The highest temperature is required because a smaller radiator is installed compared to the rooms of the same size. As described before, the hallway is assumed with a fixed low temperature. 

\begin{figure}[htbp]
\centering
  \includegraphics[width=0.7\linewidth]{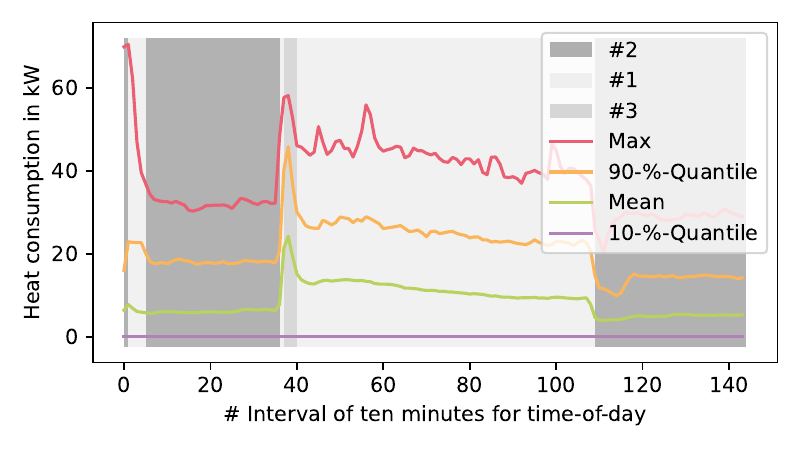}
  \caption{Quantiles of heat consumption data over ten-minute-intervals for building B.}
  \label{fig:Quantiles:buildB}
\vspace{-12pt}   
\end{figure}

For building B, the consumption data over the time-of-day shows significant variance. The maximum, 90\%-quantile, mean and 10\%-quantile per 10-minute interval are presented in Figure~\ref{fig:Quantiles:buildB}. 
The background shading indicates the assignment of the respective interval to one of the identified clusters.

\begin{figure}[htbp]
\centering
  \includegraphics[width=0.7\linewidth]{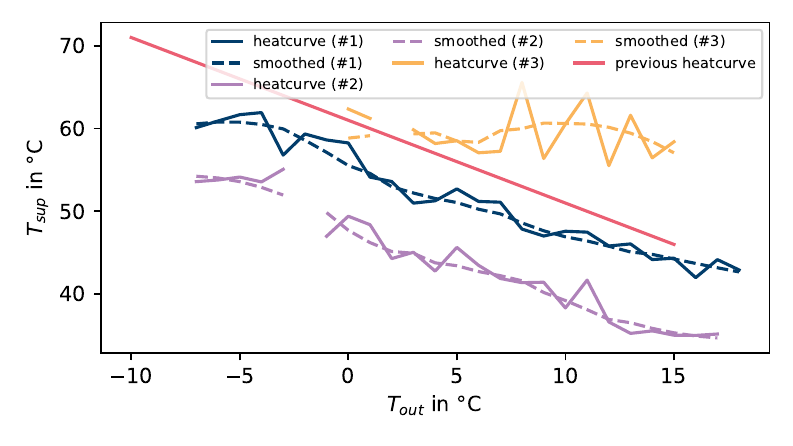}
  \caption{Heatcurve per cluster and previously active heatcurve for building B.}
  \label{fig:heatcurve:comparison:buildB}
\vspace{-12pt}   
\end{figure}

The clusters found are well interpretable: During night-times, the demand is lowered (\#2, shading in dark gray) while a peak occurs on heating up in the morning (\#3, shading in medium gray). During the day, the third cluster is identified (\#1, shading in light gray). As at some days there was a short heat-up also just after midnight, the clustering routine does also assign a short amount of time during the night to this cluster.
Based on this data analysis, the number of clusters was chosen to three.
As a result, also three different heatcurves are calculated. These heatcurves are shown in Figure~\ref{fig:heatcurve:comparison:buildB} as well as the set heatcurve during data sampling. 

It becomes visible that considering a clustering of the historical consumption data allows for more precise lowering of the supply temperature related to various heat demand characteristics. Thus, during times assigned to cluster \#1 and \#2 a significant reduction is identified: Regarding means, the cluster during night-times \#2 
is 12.99~K lower than the mean of the heatcurve during data sampling. The cluster identified for heat demand during the day \#1 
is still ca. 5.16~K lower than the previously active heatcurve.
However, for the identified cluster during heating up time span in the morning \#3, the determined heatcurve is above the set temperature during data handling. Due to the available measurement data for building B, the heatcurve of cluster \#3 shows only supply temperatures for relative high outdoor temperature above 0~°C. However, the maximal temperature is still on the same level as the heatcurve of cluster \#1 an \#2. Due to the fact that building B being young in age, the amount of measurement data available is limited. For certain outdoor temperatures the amount of available data was so small, that no relevant information could be deduced, resulting in the gaps in the heatcurves of \#2 and \#3.

\begin{figure}[htbp]
\centering
  \includegraphics[width=0.7\linewidth]{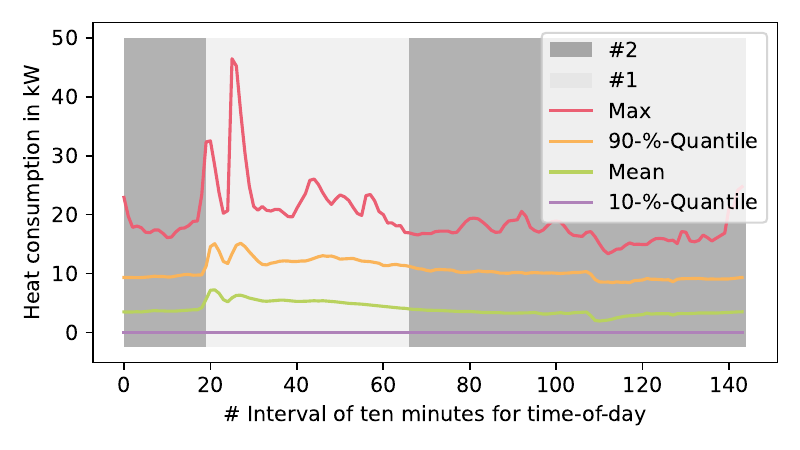}
  \caption{Quantiles of heat consumption data over ten-minute-intervals for building C.}
  \label{fig:Quantiles:buildC}
\vspace{-16pt}   
\end{figure}
\begin{figure}[htbp]
\centering
  \includegraphics[width=0.7\linewidth]{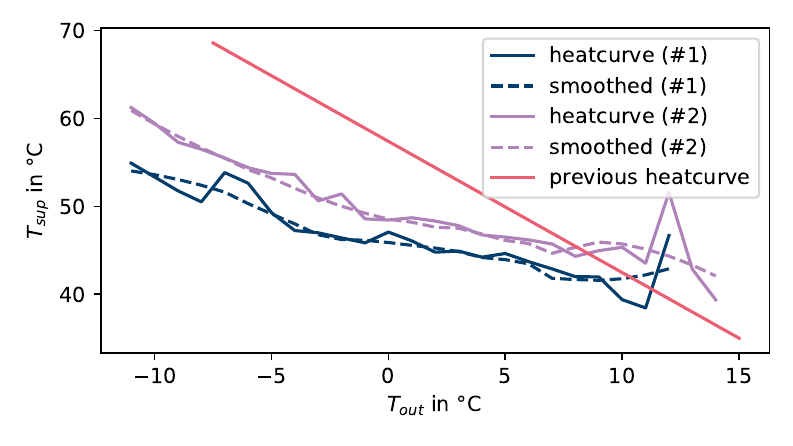}
  \caption{Heatcurve per cluster and previously active heatcurve for building C.}
  \label{fig:heatcurve:comparison:buildC}
\vspace{-12pt}   
\end{figure}

The heatcurves for building C where calculated with two different clusters. The corresponding heat consumption data and the resulting clusters are shown in Figure~\ref{fig:Quantiles:buildC}. Figure~\ref{fig:heatcurve:comparison:buildC} presents the resulting heatcurves for building C. 

Similarly to results for building B, introducing the usage of two clusters shows a lowering for the first cluster \#1 of ca. 2.7~K compared to the second cluster \#2 in the mean over the span of outdoor temperatures from -7°C to 15°C (see Figure~\ref{fig:heatcurve:comparison:buildC}). 
Regarding to the heatcurve during data sampling, the significant change in slope of the new determined heatcurves is to be underlined. Regarding means, the lowering of the heatcurves suggested by the method is about 3.49~K respectively 6.2~K. However, the temperature reduction during low outdoor temperatures is more significant while for higher outdoor temperatures no reduction or even a higher supply temperature is identified than set during data sampling.
The rise of supply-temperature as given by the resulting heatcurves at higher outdoor temperatures (between 10~°C and 15~°C) is an artefact from the input data and thereby inherent to a data-driven method. Such behaviour can be corrected by additional postprocessing. The applied filter already provides significant improvement here. 

\subsection*{Evaluation by application to real building}
\begin{figure}[htbp]
\centering
  \includegraphics[width=0.7\linewidth]{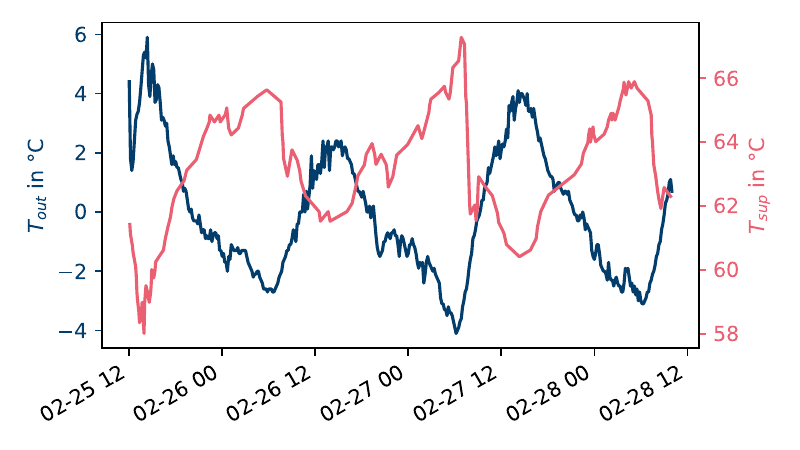}
  \caption{Timeseries data of outdoor air temperature and supply temperature setpoint during real-live experiment.}
  \label{fig:timeseries:realExperiment:buildA}
\vspace{-12pt}   
\end{figure}

For evaluation, the derived heatcurve has been applied to the real-live operation within building A for a first method validation in the scope of this study. Building A is chosen for the experiment, since the valve opening of the heaters installed in the rooms can be measured in this building. The adapted heatcurve is applied to the building over a timespan of three days.
The measured outside temperature and the set supply temperature, controlled by the adapted heatcurve, are shown in Figure~\ref{fig:timeseries:realExperiment:buildA}.

\begin{figure}[htbp]
\centering
  \includegraphics[width=0.7\linewidth]{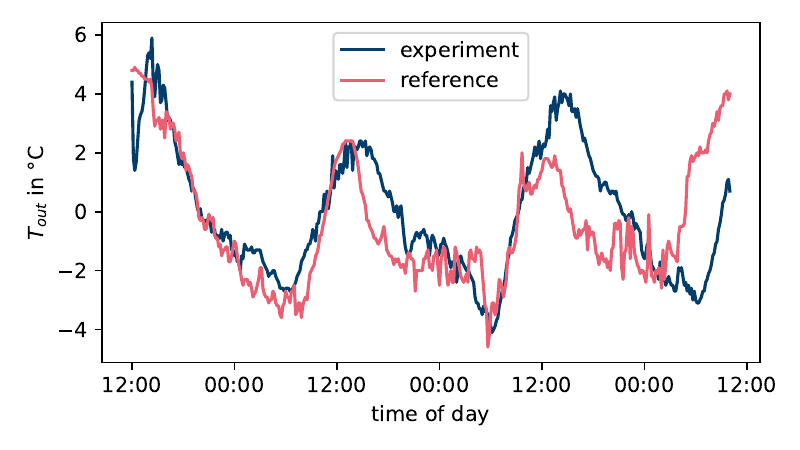}
  \caption{Comparison of outer temperature between the time of the experiment and the reference.}
  \label{fig:timeseries:realExperiment:buildA:temps}
\vspace{-12pt}   
\end{figure}

For the comparison of the adapted heatcurve to the previously applied heatcurve, measurement data from 2021 are taken into consideration. To find a comparable time horizon, the trajectory of measured outdoor temperatures during the three-day experiment (exp) are compared to the available reference (ref) data of 2021. The trajectory of the exp data is compared to the ref data over the year. The root mean square error is calculated on the temperature deviation between the two time frames. 
The trajectory within the ref data with the smallest deviation to the exp data is taken for method evaluation. 
The found trajectories of outdoor temperature with the smallest deviation, used for heatcurve evaluation, are shown in Figure~\ref{fig:timeseries:realExperiment:buildA:temps}.

\begin{figure}[htbp]
\centering
  \includegraphics[width=0.7\linewidth]{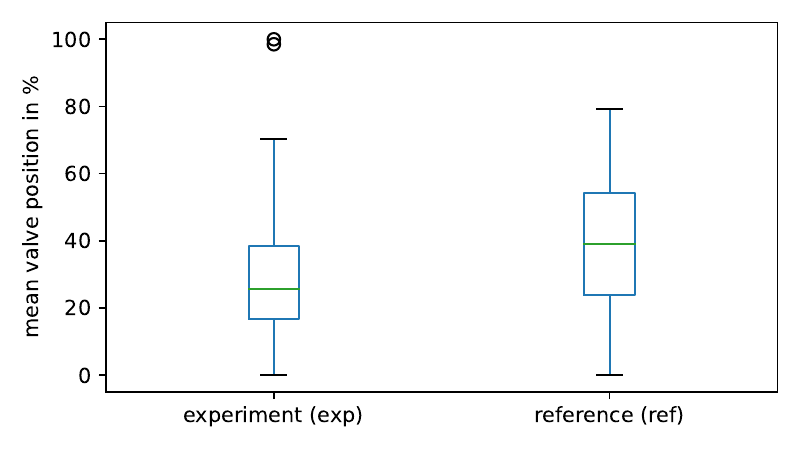}
  \caption{Boxplot on the mean valve positions per heater for exp and ref data sets.}
  \label{fig:valvePosition:Boxplot}
\vspace{-12pt}   
\end{figure}

For the evaluation of the adapted heatcurve, the effect at heater level is compared between both available measurement sets, ref and exp. The mean valve openings of the heaters in the rooms are compared, to evaluate the heater operation at the comparable outdoor temperatures. The valve opening of the heater represents the required mass flow at the heater to fulfil the desired heat supply to the room at the set supply temperature. Thus, a lower valve opening value than 100~\% indicates that the heater can supply the users' heat demand at the set temperature level.
Figure~\ref{fig:valvePosition:Boxplot} shows the boxplot of the mean valve positions over all heaters in building A.

As seen by the shown boxplots of both data sets, the mean valve opening values for the experiment with the adapted hatcurve is not higher than for the ref data, the interquartile range is even a lower. The partly lower values can be argued with different boundary conditions between the two data sets, such as the presences of occupants or solar irradiation, or by the fact that the adapted heatcurve of building A is partly higher than the previously active one (see Figure~\ref{fig:heatcurve:comparison:buildA}). 
As the mean valve opening values are quite low, it can be concluded that the requested heat demand by the users can be supplied to the room with the adapted heatcurve. 
However, there are two outliers around 100~\% in the exp data (see Figure~\ref{fig:valvePosition:Boxplot}). These two datapoints have to be evaluated further, to identify possible measurement failures or insufficient heat supply to the corresponding rooms, since the complete open valves indicate a maximum mass flow at the heater and thus a maximum heat supply. However, a maximum mass flow at the heater does not necessarily indicate an insufficient heat supply to the room. These few outliers could also result from user behaviour, e.g. higher desired indoor temperatures than assumed or continuously tipped windows.
An indicator for this assumption is the missing of complaints about comfort reduction during the real-live experiment of the adapted heatcurve. 
In the LLEC project, the installed sensors detect the presence of occupants or the state of the windows. This measurement data is used to evaluate user behaviour and is taken into account in a future extended real-live experiment for evaluation the presented method of heatcurve adaption.

\section*{Discussion}
\label{sec:Discussion}
Based on the method steps described and the results shown for several buildings, the method characteristics and application fields can be evaluated:
Due to the gray-box and data-driven character of the method, the need of appropriate input data is evident.
To allow for simple and fast calculation, the method depends on several assumptions and simplifications. These also influence the results and should be taken with care.
The room load calculation, which is based on the building heat consumption data and the heat transfer coefficient values of typical building types, need to be evaluated by measurements at room level. Measuring the heat supply at the heaters, e.g. with heat allocators as done by \citet{Benakopoulos.2022}, the approach of heat load estimation can be evaluated against real data. Thus, the approach of room heat load estimation could be assessed further, which enhances the comparison to the available heater capacity and leads to a more accurate estimation of the possible heatcurve reduction for the entire building. 

As demonstrated at the three example buildings, the method leads to plausible results. The limited required data input and adaption effort of the described method enable a quick estimation of reduction potentials in existing buildings and, thus, an efficiency exploitation with limited expenditures in an district. 
The presented evaluation experiment could only be conducted about a limited time span of three days and the evaluation possibilities were also limited in the scope of this study.
However, the heatcurve application at a real building shows promising results that the determined lowered heatcurves do not lead to any serious comfort reductions for the occupants. 
Thus, this evaluation indicates that the method leads to suitable  heatcurves, which are reasonable compared to the previous applied heatcurves, but still indicate a possible temperature reduction, especially at the lowest outside temperatures.

In future work an extended evaluation of the determined heatcurves in real applications should be carried out, for further buildings but also in the context of a long term evaluation, in order to consider various building characteristics and different outdoor temperatures.
Moreover, our study focuses on radiator-based heating systems. An extension to air-conditioning systems would further enlarge the range of buildings the method is applicable to.

\section*{Conclusion}
\label{sec:Conclusion}
Reduction of supply temperatures in buildings has multiple positive effects on different heat supply systems, such as heat pumps or district heating. Adjusting the heatcurve that controls the supply temperature of heating systems in existing buildings is an easy and effective measure. However, the determination of possible supply temperature reduction is currently determined by detailed simulation models or installed heat allocators on room level. Thus, the expenditure is high to utilize these approaches in several buildings.
To exploit the potential of lower supply temperatures in existing buildings on a district level, we developed an approach to estimate the actual temperatures required by a heating system based on historical building heat consumption data and information about the heater installed in the rooms. 

The presented method determines the required supply temperature for each room by calculating the room heat load with the overall building heat consumption and comparing it with the installed heater capacity in each room. Thus, the actual required temperature for each heater is specified. The supply temperature at the building level could be adjusted by implementing an adapted heatcurve at the central heating system.
In this way, we estimate the potential for lowered supply temperatures with minimal input data, limiting the effort to apply the method to several buildings in a district.
The resulting heatcurves can be applied easily as parametrisation to the already existing automation. Such non-invasive measure results in low implementation costs.
For further lowering of the supply temperature, the critical heaters can be identified from intermediate results directly (see Figure~\ref{fig:floor_plan:buildA}) allowing an effective use of financial effort when measures should be extended to invasive measures.

The method is applied to three exemplary buildings with various characteristics and leads to useful results of adapted heatcurves. Furthermore, one of the adapted heatcurve is applied to a building in real operation. For the evaluation time span, we could conclude a sufficient heat supply to the rooms and user comfort.
In future work, additionally determined heatcurves have to be applied to several buildings in order to evaluate the method for different types of buildings and identify possible improvements for the individual steps of the method.

\section*{Acknowledgement}
We gratefully acknowledge financial support by BMWK (German Federal
Ministry of Economic Affairs and Climate Action), grant number 03ET1551A.
Moreover, the collaboration with the colleagues at IEK-10, at the infrastructure departments and project management at FZJ is highly appreciated.

\bibliographystyle{IEEEtranN}
\bibliography{BS2023_literature}
\newpage

\end{document}